\documentstyle[aps,prl]{revtex}

\begin{document}
\input psfig

\begin{titlepage}

\vspace{\fill}

\title{THE PATH INTEGRAL FOR THE LOOP REPRESENTATION 
OF LATTICE GAUGE THEORIES}

\author{J. M. Aroca}
\address{Departament de Matem\`atiques, Universitat
Polit\`ecnica de Catalunya
, Gran Capit\`a, s/n Mod C-3 Campus Nord,
08034 Barcelona, Spain.}

\author{H. Fort and R. Gambini}
\address{Instituto de F\'{\i}sica, Facultad de Ciencias, 
Tristan Narvaja 1674
,11200 Montevideo, Uruguay}


\date{\today}
\maketitle
\date{\today}

\nopagebreak

\begin{abstract}
We show how the Hamiltonian lattice 
$loop$ $representation$ can be cast 
straightforwardly in the path integral formalism. 
The procedure is
general for any gauge theory. Here we present in detail the simplest 
case: pure compact QED. 
The lattice loop path integral approach allows to knit together
the power of statistical algorithms with the transparency of the
gauge invariant loop description. 
The results produced by numerical simulations
with the loop classical action for different lattice models
are discused.
We also analyze the lattice path integral in terms of loops for the
non-Abelian theory. 
\end{abstract}

\end{titlepage}

\section{Introduction}

	The loop approach to abelian quantum 
gauge theories was introduced 
in the early eighties \cite{gt}. Later it was generalized
to the non-abelian Yang-Mills gauge theory \cite{gt2}. This
Hamiltonian method allows to formulate 
gauge theories in terms of their natural
physical excitations: the loops.
The original aim of this general description
of gauge theories was to avoid gauge redundancy 
working directly in the space of the
gauge invariant excitations. However, soon it 
was realized that the loop
formalism goes far beyond of a simple 
gauge invariant description.
The introduction by Ashtekar \cite{a} 
of a new set of variables
that cast general relativity in the same language as 
gauge theories allowed to apply loop techniques as a 
natural non-perturbative description of Einstein's theory.
In particular, the loop representation appeared as the most
appealing application of the loop techniques to this problem
\cite{rs}. The covariant worldsheet formulation of quantum gravity  
corresponding to the canonical 
loop representation is still unknow. This 
version of loop quantization would make the general 4-dimensional 
diffeomorphism invariance manifest 
and it will be probable more suitable 
to tackle some crucial problems as the black hole dynamics. Some 
progress in that direction has been reported recently \cite{rei}.

\vspace{3mm}

The Hamiltonian analytical computation techniques for gauge
theories have been
developed during the last decade and they provide qualitatively
good results for several lattice models \cite{glt}-\cite{af}. 
On the other hand  
a Lagrangian approach in terms of loops has been elusive, due mainly
to the non-canonical character of the loop algebra. 
This feature forbids
the possibility of performing a Legendre transformation as 
a straightforward way to obtain the Lagrangian from the Hamiltonian. 
A path integral loop formulation will allow to employ 
the more powerful statistical computation
techniques. Gauge invariant actions 
corresponding to the Villain form of 
the U(1) model were proposed recently \cite{abf}
and generalized to 
include matter fields \cite{abfs}. 
       
        These actions has been used as a computational tool.
A  Metropolis Monte-Carlo algorithm was implemented fixing
the acceptance ratio, as it is usual in random surfaces analysis.
The U(1) model was studied for different
lattice sizes \cite{abf}, 
imposing periodic boundary conditions. Simple thermal 
cycles showed the presence of a phase transition in the
neighborhood of $\beta_V=0.639$. 
In the case of
of U(1)-Higgs model,
the $\beta$-$\gamma$ phase diagram was mapped out \cite{abfs}. 
In both models, by virtue of the gauge invariance
of this description, the equilibrium configurations 
were reached faster
than with the ordinary gauge-variant descriptions.
Furthermore, the absence
of the typical strong metastability 
of Monte Carlo analysis
, which makes very difficult the
numerical analysis of the phase transition critical exponents,
was observed. This fact is
one of the main advantages for using the loop action.

In this paper we want to address the issue of the 
explicit correspondence between these actions and the Hamiltonian
loop representation using the transfer matrix techniques.

\vspace{3mm}

This paper is organized as follows. In section 2 we show how the 
loop description, originally introduced in  
the Hamiltonian formalism, can be cast
in the lattice path integral formalism. We illustrate this 
in detail for the Abelian case (compact electromagnetism).
The path integral of lattice $U(1)$ theory is expressed 
as a sum of the world sheets of electric loops.
We discuss the connection of this classical loop action with
the Nambu string action. 
In section 3 we consider the extension to the path integral
loop formalism to the case of non-Abelian Yang-Mills fields.
In section 4 we conclude with some remarks.

\vspace{5 mm}

\section{The lattice path integral in terms 
of loops: electromagnetism.}

The loop based approach of ref.\cite{gt} describes the quantum
electrodynamics in terms of the gauge invariant holonomy
(Wilson loop)
\begin{equation}
\hat{W} (\gamma ) = \exp [i e \oint_{\gamma} A_a (y) dy^a] , 
\label{eq:Wloop}
\end{equation}
where $\gamma$ is a spatial loop at constant time $t$.
$\hat{E}^a (x)$ is the  conjugate  electric  field operator. 
They  obey   the
commutation relations
\begin{equation}
[\hat{E}^a(x), \hat{W} (\gamma)] = e \int_\gamma 
\delta(x - y) dy^a \hat{W}
(\gamma) .
\label{eq:alg}
\end{equation}

     These  operators  act  on  a  state  space  of  abelian  loops
$\psi(\gamma)$ that may be expressed in terms of the transform
\begin{equation}
\psi (\gamma) = \int d_\mu [A] <\gamma \mid A> <A \mid \psi>
= \int d_\mu [A] \psi [A] \exp [- i e \oint_{\gamma} A_a dy^a] .
\label{eq:trans}
\end{equation}

This loop representation has many appealing features: i)
It allows to do away with the first class constraints 
of gauge theories
i.e. the Gauss law is automatically satisfied. ii) The 
formalism only involves gauge invariant objects i.e. no
gauge redundancy. iii) All the
gauge invariant operators have a transparent geometrical meaning 
when they are realized in the loop space.

When this loop representation is implemented in the lattice 
equation (\ref{eq:alg}) becomes
\begin{equation}
[\hat{E}_{\ell},\hat{W}(\gamma)] =  N_{\ell}(\gamma)\hat{W}(\gamma),
\label{eq:alg1}
\end{equation}
where $\ell$ denotes the links of the lattice, $\hat{E}_{\ell}$
the lattice electric field operator, 
$\hat{W}(\gamma)=\prod_{\ell \in \gamma}\hat{U}(\ell)$ and 
$N_{\ell}(\gamma)$ is the number of times that the link $\ell$
appears in the closed path $\gamma$.

In this loop representation, the Wilson loop 
acts as the loop creation operator:
\begin{equation}
\hat{W}(\gamma')\mid \gamma> = \mid \gamma'\cdot \gamma>.
\label{eq:Wloop1}
\end{equation}

     The physical meaning of an abelian loop may be deduced  from
(\ref{eq:alg1}) and (\ref{eq:Wloop1}), in fact
\begin{equation}
\hat{E}_{\ell} \mid \gamma> = N_{\ell}(\gamma) \mid \gamma>,
\label{eq:E}
\end{equation}
which implies  that  $\mid \gamma>$  is  an  eigenstate  of  the
electric field. The corresponding eigenvalue is different from 
zero if the link $l$ belongs to
$\gamma$. Thus $\gamma$ represents a confined  line  
of  electric flux. 

     The $U(1)$ lattice Hamiltonian can be written
in terms of both previous fundamental operators as
\begin{equation}
\hat{H}=\frac{g^2}{2} 
\sum_l \hat{E}_\ell^2 
-\frac{1}{2g^2}\sum_p (\hat{W}_p+\hat{W}_p^\dagger).
\label{eq:H}
\end{equation}

\vspace{5mm}

        Now, starting from the path integral for the Wilson 
U(1) lattice action, let us show how to set up the loop path
integral.
\begin{equation}
Z_W = \int_{-\pi}^{\pi} [d\theta_{\ell} ]\exp(\frac{\beta}{2}
\sum_p \cos 
{\theta}_p),
\label{eq:ZWA}
\end{equation}
where the subscripts $\ell$ and $p$ stand for the lattice links 
and plaquettes respectively, $\beta=\frac{1}{e^2}$ ($e$ is the
electric charge of the electron) and $\theta_\ell$ is a
compact variable  $\in [-\pi,\pi]$ 
attached to links and  
$\theta_p$ is its "discrete curl".
Fourier expanding the $\exp [ \beta/2 \cos \theta ]$ we get
\begin{equation}
Z_W = \int_{-\pi}^{\pi} [d\theta_{\ell} ]  \prod_p \sum_{n_p} 
I_{n_p}(\beta ) 
e^{in_p {\theta}_p},
\end{equation}
where the $I_n$ are modified Bessel functions.
$Z_W$ can be written as
\begin{equation}
Z_W = \sum_{ \left\{ n_p \right\} }
\int_{-\pi}^{\pi} [d\theta_{\ell} ] \exp 
(\sum_p \ln I_{n_p}(\beta ) ) 
e^{i<n , \nabla \theta_{\ell} >},
\end{equation}
where we use the notations of the calculus of 
differential forms on the lattice.
In the above expression:  $\theta$ is a
real compact 1-form defined in each link of the lattice 
$\nabla$ is the co-boundary operator --which maps $k$-forms
into $(k+1)$-forms--, $n_p$ are integer 2-forms defined at 
the lattice plaquettes and consider the scalar product of p-forms 
$<f \mid g> = \sum_{c_k} f(c) g(c)$ where the sum runs
over the $k$-cells $c_k$ of the lattice 
($c_0$ sites, $c_1$ links and so on ).
Under this product the $\nabla$
operator is adjoint to the border operator $\delta$ which maps
k-forms onto (k-1)-forms and which 
corresponds to minus times the usual
divergence operator i.e.
\begin{eqnarray}
<f \mid  \nabla g> = <\delta f \mid g>,\\
<\nabla f \mid g> = <f \mid \delta g>.
\label{eq:inter}
\end{eqnarray}
Using equation (\ref{eq:inter}) and integrating over $\theta_{\ell}$
we get a $\delta (\delta n_p)$.
Then, we arrive to
the following functional of the integer 2-forms $n$
\begin{equation}
Z_W[n] \propto \sum_{ \left\{ n_p; \delta n_p = 0 \right\} }
\exp (\sum_p \ln I_{n_p}(\beta ) ), 
\label{eq:ZL1}
\end{equation}
the constraint  $\delta n_p = 0$ means that 
the sum is restricted to
$closed$ 2-forms. Thus, the sum runs over 
collections of plaquettes
constituting closed surfaces.

An alternative and more easy to handle 
lattice action than the Wilson
form is the Villain form. The partition 
function of that form 
is given by
\begin{equation}
Z_V = \int [d\theta ] \sum_{ s }\exp
(-\frac{{\beta}_V}{2}\mid\mid \nabla \theta -2\pi s\mid\mid^2),
\label{eq:Villain}
\end{equation}
where $\mid \mid \ldots \mid \mid^2 = <\ldots , \ldots>$.
If we use the Poisson summation formula
$$\sum_s f(s) = \sum_n \int_{-\infty}^{\infty} d\phi 
f(\phi) e^{2\pi i\phi n}$$
and we integrate the continuum $\phi$ variables we get
\begin{equation}
Z_V = (2\pi {\beta}_V)^{-N_p/2} \int [d\theta ] \sum_{n }\exp
(-\frac{1}{2{\beta}_V}<n,n>+i<n,\nabla \theta >),
\label{}
\end{equation}
where $N_p$ in the number of plaquettes of the lattice. Again, we can use
the equality:
$<n,\nabla \theta>=<\delta n,\theta >$ and integrating over $\theta$
we obtain a $\delta (\delta n)$. Then, we get 
\begin{equation}
Z_V [n]= (2\pi {\beta}_V)^{-N_p/2}  
\sum_{\left\{ n; \delta n = 0 \right\}}
\exp(-\frac{1}{2{\beta}_V}<n,n>).
\label{eq:ZL2}
\end{equation}

\begin{center}
\begin{figure}[t]
\hskip 1cm \psfig{figure=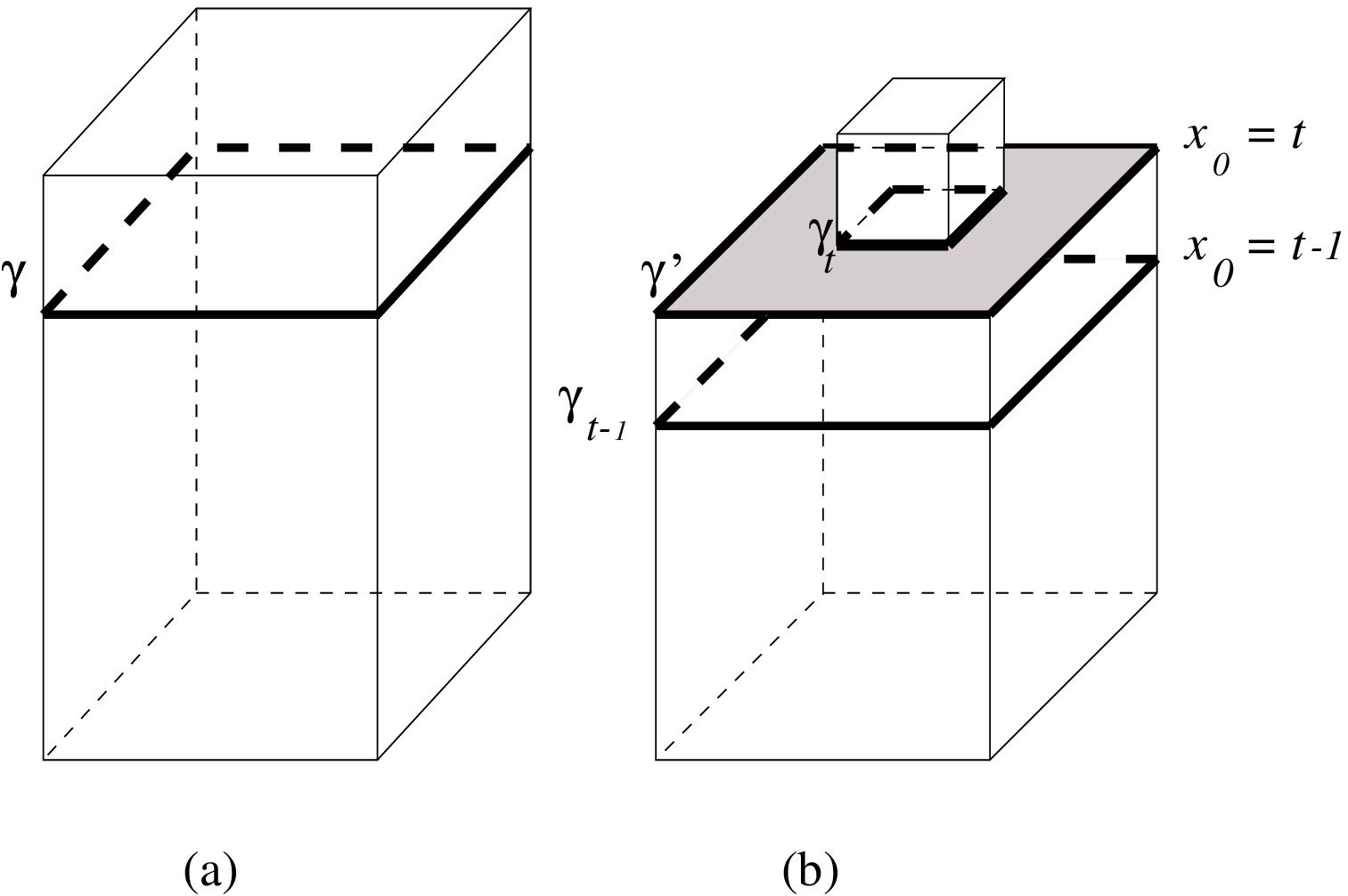,height=7cm}
\caption{}
\label{figI}
\end{figure}
\end{center}

Let us first discuss the (2+1) dimensional case.
In figure \ref{figI} we show the result 
of intersecting  one of the previous surfaces with a 
$t=\mbox{constant}$ plane. We can get: i) a spatial closed 
path or loop  (Figure  1-(a) )
or ii) an open surface connecting
the loop $\gamma_t'$, obtained by translating the 
loop $\gamma_{t-1}$ by one temporal lattice unit,
and $\gamma_{t}$ (the shaded surface in Figure 1-(b) ).
A loop $\gamma_t$ living 
on the $t$ slice is properly specified by the temporal plaquettes
which leave this slice. This is equivalent to
say that in two spatial dimensions,
given the loops at time $t$ and time $t+a_0$, the surface
which connect them is unambiguously defined.

    In more than 3 space-time dimensions
the situation is different, the loops $\gamma_{t-1}$
and $\gamma_t$ do not define a unique world sheet
connecting them. Thus,
if we consider the intersection of one of the world-surfaces
with a $t=\mbox{constant}$ plane we can get:
i) a loop $\gamma_t$ ii) an open surface connecting
the loop $\gamma_t'$ and $\gamma_{t}$ 
iii) a closed spatial surface.
The situation is completely analogous
to the path integral of a free particle on a (2+1)
dimensional lattice 
i.e. the world lines connecting  points at different 
times intersected with a $t=\mbox{constant}$
plane give: i) points or 
ii) open paths connecting the intersecting
point of the timelike link at 
$p_{t-1}$ with surface $t=\mbox{constant}$
and the point $p_t$ and iii) loops associated  
with different
choices of the previous mentioned open paths.

Now we will show that the expression
(\ref{eq:ZL2}) 
is a possible "loop path integral" $Z_L$ 
\footnote{Another possible 
path integral in terms of loops is given by 
(\ref{eq:ZL1}).}. First, it is easy
to prove that the creation operator of these loops is just
the creation operator of the loop representation, namely 
the Wilson loop operator. Repeating the steps from 
equation (\ref{eq:Villain}) to equation (\ref{eq:ZL2}) we get for
$<\hat{W}(\gamma_t)>$
\begin{equation}
<W(\gamma_t)> = \frac{1}{Z}(2\pi {\beta}_V)^{-N_p/2}  
\sum_{
                \begin{array} {c} n\\
                              \left(  \delta n = \gamma_t \right)
                \end{array}
                 }\exp
(-\frac{1}{2{\beta}_V}<n,n>).
\label{eq:W}
\end{equation} 
This is a sum over all world sheets
spanned on the loop $\gamma_t$.
Secondly, by means of the transfer matrix 
method let us show that we
re-obtain the Hamiltonian (\ref{eq:H}) from (\ref{eq:ZL2}).
As we wish to consider the continuous time limit of
the previous lattice Euclidean space-time theory, we
introduce a different lattice spacing $a_0$ 
for the time direction.
The couplings on timelike and spacelike
plaquettes are no longer equal in the action
i.e. we have two coupling constants: $\beta_0$ and 
$\beta_s$. The temporal coupling constant $\beta_0$ 
decreases with $a_0$ whilst the spatial coupling constant
$\beta_s$ increases with $a_0$.
We wish to find an operator $\hat{T}$ 
over the Hilbert space of loops $\{ |\gamma > \}$
such that, the loop path integral $Z_L$ is given by

$$Z_L=\prod_t <\gamma_{t+a_0}| \hat{T} |\gamma_t>.$$

$\hat{T}$ is related, when $a_0$ is small,
with the Hamiltonian $\hat{H}$ by
\begin{equation}
\hat{T}\propto e^{-a_0\hat{H} + O(a_0^2)}.
\label{eq:T-H}
\end{equation}
The transfer matrix 
between times $t$ and $t+a_0$
in the loop representation of kets $|\gamma>$ can be written as
\begin{equation}
<\gamma_{t+a_0} | \hat{T} |\gamma_t>=\exp{ [ \, \frac{1}{2\beta_0}
\sum_{p_t} n_p^2 - \frac{1}{2\beta_s}
\sum_{p_s \in (S_{t+a_0}-S_t)} n_p^2 \, \, ]} \, ,
\label{eq:Tm}
\end{equation}
where $S_{t+a_0}$ and $S_t$ are the surfaces enclosed by the
loops $\gamma_{t+a_0}$ and $\gamma_t$ respectively.
Due to the fact that the surfaces are closed
the integers $n_p$ of the temporal plaquettes which
depart from the loop $\gamma_t$ at time $t$ and 
arrive to the loop  $\gamma_{t+a_0}$ at time $t+a_0$ 
are equal to the number of times the spatial link $\ell_t$
appears in the loop $\gamma_t$, $N_\ell$.
Therefore, the equation (\ref{eq:Tm}) can be re-written as  
\begin{equation}
<\gamma_{t+a_0} | \hat{T} |\gamma_t>=\exp{ [ \,-\frac{1}{2\beta_0}
\sum_{\ell} N_\ell^2(\gamma_t) - \frac{1}{2\beta_s}
\sum_{p_s \in (S_{t+a_0}-S_t) }  n_p^2 \, ]}.
\label{eq:Tm2}
\end{equation}
The kets $|\gamma_{t+a_0}>$ and $\gamma_t>$ are connected by 
\begin{equation}
|\gamma_{t+a_0}>= \prod_{p \in \Delta_0 S_\gamma} \hat{W}_p^{n_p}
|\gamma_t>.
\label{eq:gammas}
\end{equation}
Using equations (\ref{eq:Wloop1}), (\ref{eq:E}) 
and (\ref{eq:gammas})
(\ref{eq:betas}) we get 
\begin{equation}
\hat{T} = \sum_{ \{ \Delta_0 S_\gamma \} }
\prod_{p \in \Delta_0 S_\gamma} \hat{W}_p^{n_p}
\exp{  [ \, -\frac{1}{2\beta_0} \sum_\ell \hat{E}_\ell^2 
-\frac{1}{2\beta_s}
\sum_{p_s \in (S_{t+a_0}-S_t) } n_p^2  \, ] \, },
\label{eq:T}
\end{equation}
where $\{ \Delta_0 S_\gamma \}$ are all the possible
modifications between the surfaces enclosed by the loops 
$\gamma_{t+a_0}$ and $\gamma_t$ i.e. it denotes  
the set of possible configurations $\{ n_p \}$
of integers attached to the plaquettes 
$p \in \Delta_0 S_\gamma$ . Then,
\begin{equation}
\hat{T} = \prod_p \sum_{n_p}
[ \, \hat{W}_p^{n_p}
\exp{ ( \, -\frac{1}{2\beta_s}
\sum_{p_s \in (S_{t+a_0}-S_t) } n_p^2  \, ) } \, ] 
\exp{  [ \, -\frac{1}{2\beta_0} \sum_\ell \hat{E}_\ell^2 \, ] }.
\label{eq:T2}
\end{equation}
To obtain a proper classical limit we should take 
\begin{eqnarray}
\beta_0= \frac{a}{g^2a_0} \\
\beta_s= \frac{1}{2} 
\frac{1}{\mbox{ln} (2g^2 a/a_0)} \, ,
\label{eq:betas}
\end{eqnarray}
where $a$ continues to denote the spacelike spacing.
This implies that for $a_0$ small
the operator $\hat{T}$ is given by
\begin{equation}
\hat{T} =
\exp \{ -a_0 [ \, \frac{g^2}{2a} \sum_\ell \hat{E}_\ell^2
+ \frac{1}{2ag^2} \sum_p (\hat{W}_p+\hat{W}_p^\dagger) \, ] 
\, + O(a_0^2) \} ,
\label{eq:T3}
\end{equation}
i.e. we recover the Hamiltonian (\ref{eq:H}).
This definitely confirms that
the expression of the partition 
function of compact electrodynamics in terms of the world 
sheets of loops: the $loop$ (Lagrangian) representation.

        From equation 
(\ref{eq:ZL2}) we can observe that the loop action 
is proportional to the $quadratic$ $area$ $A_2$:
\begin{equation}
S_L = -\frac{1}{{\beta}_V} A_2 = 
-\frac{1}{{\beta}_V}\sum_{p \in {\cal S}} n_p^2 = 
-\frac{1}{{\beta}_V}<n,n>,
\label{eq:A2}
\end{equation}
i.e. the sum of the squares 
of the mul\-ti\-pli\-ci\-ties $n_p$ of pla\-que\-ttes which 
constitute the loop's
world sheet { \cal S}. It is interesting to note the similarity  
of this action with the continuous Nambu action 
or its lattice version, the Weingarten action \cite{we}
which are proportional to the area swept 
out by the bosonic string 
\footnote{The relation between the surfaces 
of the Wilson action and 
those of Weingarten action has been analyzed by Kazakov et al
in ref.\cite{kkm}.}.

\section{The Non-Abelian Loop Action}

   Let us see how can be extended the 
path integral loop description for the non-Abelian case 
of Yang-Mills theory. The path integral for the Wilson 
action for a general non-Abelian compact gauge group G is given by
\begin{equation}
Z_W = \int [d U_{\ell} ]\exp [{\beta\sum_p \mbox{Re} 
( \mbox{tr} \, U_p ) \,]}, 
\label{eq:ZWNA}
\end{equation}
where the $U_\ell \in G$ and $U_p = \prod_{\ell \in p} U_\ell$.
Equation (\ref{eq:ZWNA}) reduces to (\ref{eq:ZWA})
for the case G $\equiv$  U(1).
The analogous of the Fourier expansion
for the non-Abelian case is the {\em character} expansion.
The characters $\chi_r(U)$ of the
irreducible (unitary) representation $r$ of dimension
$d_r$, defined as the traces of these
representations, are an orthonormal basis for the
{\em class} functions of the group i.e. \cite{id}
\begin{eqnarray}
\int d U \chi_r(U) \chi_s^*(U)=\delta_{rs} 
\label{eq:ort1} \\
\sum_r d_r \chi_r(U{V^{-1}})=\delta(U,V).
\label{eq:ort2}
\end{eqnarray}
In particular, as a useful consequence we have 
\begin{equation}
d_r \int d U \chi_s(U) \chi_r(U{V^{-1}})=\delta_{rs}\chi_r(V).
\label{eq:ort3}
\end{equation}

By means of the character expansion we can express
\begin{equation}
\exp  \{ \beta\sum_p \mbox{Re} [ \chi ( U_p ) \,] \}
= \prod_p \sum_r c_r \chi_r (U_p),
\end{equation}
with
\begin{equation}
c_r=\int dU \chi_r^*(U)\exp{(\beta \chi(U)\, )}.
\label{eq:coef}
\end{equation}

For instance, in the case of G $\equiv$ SU(2)
the gauge fields can be parametrized as

$$U=\cos \frac{1}{2}\theta + i \sigma_a n_a \sin \frac{1}{2}\theta,
\;\;\;\;\;\;\; 0 \le \theta \le 4\pi \, ,$$
 
and the corresponding
irreducible representations are classified by a non-negative
integer or half-integer spin $j$
i.e. $r\equiv j$ and the characters are given by
\begin{equation}
\chi_j(U)= \frac{\sin (j+\frac{1}{2})\theta}
{\sin \frac{1}{2} \theta}.
\label{eq:ch}
\end{equation}
A direct application of (\ref{eq:coef}) yields the $c_j$
in terms of modified Bessel functions, and therefore
we can express (\ref{eq:ZWNA}) as 
\begin{eqnarray}
Z_W = \int [d \theta_{\ell} ] \prod_p 
[ \sum_{j_p} 2(2j_p+1)\frac{I_{2j_p+1}(\beta)}{\beta}
\frac{\sin (j+\frac{1}{2})\theta_p}
{\sin \frac{ \theta_p}{2} } ] 
\nonumber \\
=\sum_{ \{ j_p \} } \prod_p [ 2(2j_p+1)
\frac{I_{2j_p+1}(\beta)}{\beta} ] \int [dU_\ell] \prod_p
\frac{\sin (j_p+\frac{1}{2})\theta_p}
{\sin \frac{\theta_p}{2}}.  
\label{eq:ZWNA2}
\end{eqnarray}
A given subset of plaquettes is homeomorphic to a simple
surface if any link bounds at most two plaquettes
of this subset. The links bounding exactly one plaquette
make up the boundary of this surface. 
Any configuration can be
decomposed as a set of maximal simple surfaces 
by cutting it along
the links bounding more than two plaquettes.
In principle, there are 
two possibilities for the boundary curves:
either a free boundary, bounding only one simple surface or
a singular branch line along which more than two simple
surfaces meet. In fact, relation (\ref{eq:ort1}) forbids the 
existence of free boundaries for non trivial
configurations contributing to the path integral.

The integration over the internal links of the simple surfaces
is performed using (\ref{eq:ort3}). Note that the plaquettes
of a simple surface component should carry the same  
group representation. After integrating over all the inner links
of the simple components one gets an expression
involving only the links of the boundary i.e. something
proportional to

$$\prod_{\mbox{boundaries}} \chi_r (U_{\mbox{boundary}}).$$

What follows is the integration of gauge fields along the singular
branches which gives rise to the Clebsch-Gordan coefficients
coupling the different representations of the considered
gauge group G. For instance, imagine that there is
only one singular closed branch line which is
the common boundary shared by $n$ simple surfaces
with representations $r_1, r_2, ...,r_n$. The integration
over gauge fields produces a factor $N_{r_1,...,r_n}$,
which counts the number of times the trivial
representation is contained in the product
$r_1\otimes r_2 \otimes ...\otimes r_p$.

Cases in which different 
singular branch lines meet in a point are considered in 
ref. \cite{rei}. Each point of intersection
involves a Racah-Wigner symbol. Thus, one can see that the
Hamiltonian formulation associated to this action will be 
given in terms of a spin network  of colored loops \cite{penrose}
\cite{rs2}. A rigorous proof of this fact using the transfer
matrix technique is still required. This is not straightforward
because one still need to develop the lattice
Hamiltonian formulation
of Yang-Mills theory in terms of spin networks.

We can also generalize the Villain form of the action 
for any gauge group using the {\em heat kernel} action 
\cite{mo},\cite{dz}:
\begin{equation}
\exp (\beta S_{HK})=\prod_p \sum_r d_r \chi_r(U_p) \exp
[-C_rª{(2)}/N\beta]
\label{eq:HK}
\end{equation}
where $C_rª{(2)}$ is the quadratic Casimir invariant for the
representation $r$.
For G $\equiv$ SU(2) the heat kernel action reads
\begin{equation}
\exp (\beta S_{HK})=
\prod_p \sum_{j=0,1/2,..} (2j+1) 
\frac{\sin (j+\frac{1}{2})\theta_p}{\sin \frac{\theta_p}{2}
} \exp [ -j(j+1)/2\beta \,] \, ,
\label{eq:hksu(2)}
\end{equation}
while only integers values of $j$ are used for the SO(3) group.

\section{Conclusions}

As it was mentioned, the loop space provides a common scenario for a 
non-local description of gauge theories and quantum gravity.
The loop approach is no more exclusively Hamiltonian
, its Lagrangian counterpart is now available. A path integral
action for the Yang-Mills theory in terms of loop variables 
is very valuable because it combines the geometrical transparency
and economy of the loop description with the versatility to 
perform calculus. 
We have presented the state of the art in that program, which still
is an open issue.

	The path integral approach to quantum gravity 
has very appealing features. 
In particular it may provide a more suitable framework for the
development of useful approximation schemes for the
study of black hole physics and  it may allow to 
analyze issues such as the
computation of the probabilities 
for a change of the spatial topology
that seem to be very difficult to 
formulate in the canonical approach.
Even though the the connection of the 
canonical loop representation of
quantum gravity and the path integral approach is 
still an open problem
the determination of the explicit form of 
the loop actions in gauge
theories is an important step in this direction. 
An important remark is
that the lattice framework seem to be unavoidable 
in order to have well
defined loop actions. In fact in spite 
of the similarities with the
Nambu actions, the loop actions for gauge 
theories involve cuadratic
surface elements that are not well defined 
in the continuum.

        Finally, concerning the lattice loop action
as a computational tool,
we already mentioned that the results produced by 
numerical simulations
for different models 
are very encouragging.
In ref. \cite{abf},
the loop action (\ref{eq:A2}) corresponding 
to Villain form of $U(1)$ model was considered. 
The extension 
of the Lagrangian $loop$ description
in such a way to include matter fields
was also simulated \cite{abfs}. 
The lattice path integral
of $U(1)$-Higgs model is expressed as a sum over closed as much
as open surfaces.  These surfaces correspond to world 
sheets of loop-like pure
electric flux excitations and open electric 
flux tubes carrying matter fields
at their ends.  This representation is connected 
y a duality transformation
with the $topological$ representation of the 
path integral (in terms of
world sheets of Nielsen-Olesen strings \cite{no}
both closed and open connecting pairs of
magnetic monopoles).  Simulating numerically the loop action
corresponding to the Villain form, the two-coupling 
phase diagram 
of this model was mapped out. 
The gauge invariance
of the loop description bears the advantages
of economy in computational time and 
the absence of the strong metastability previously 
observed in the ordinary Monte Carlo analysis.

\vspace{3mm}
{\large \bf Acknowledgements}
\vspace{3mm}

This work has been supported in part by the CONICYT 
projects {\bf 49} and  {\bf 318}.

We wish to thank helpful comments from Abhay Ashtekar,
Jorge Pullin and Mike Reisenberger.

\end{document}